\documentclass[prd,twocolumn,showpcs,amsmath,amssymb,nofootinbib,preprintnumbers,balancelastpage,longbibliography]{revtex4-1}
\usepackage{epsfig}
\usepackage{amsmath}
\usepackage{bm}
\usepackage{times}
\usepackage{physics}
\usepackage{graphicx}
\usepackage{color}
\usepackage{slashed}
\usepackage{graphicx}
\usepackage{amsmath}
\usepackage{mathrsfs}
\usepackage{tikz}
\usetikzlibrary{positioning,shapes}
\usepackage{relsize,caption} 
\usepackage[latin1]{inputenc}
\usepackage{textcomp}
\usepackage[T1]{fontenc}
\usepackage{tikz}
\usetikzlibrary{trees}
\usetikzlibrary{decorations.pathmorphing}
\usetikzlibrary{decorations.markings}
\usetikzlibrary{positioning,arrows}
\usetikzlibrary{decorations.pathmorphing}
\usetikzlibrary{decorations.markings}
\usetikzlibrary{decorations.pathreplacing,calc}
\usetikzlibrary{decorations.pathmorphing,decorations.markings,trees,positioning,arrows}   
\newif\ifmirrorsemicircle

\def\bea{\begin{eqnarray}}
\def\eea{\end{eqnarray}}
\def\bean{\begin{equation*}}
\def\eean{\end{equation*}}

\def\bea{\begin{eqnarray}}
\def\eea{\end{eqnarray}}
\def\bean{\begin{equation*}}
\def\eean{\end{equation*}}

\begin{document}

\preprint{UCI-HEP-TR-2017-15}

\title{Millicharged Scalar Fields, Massive Photons and the Breaking of $SU(3)_{C} \times U(1)_{\rm EM}$}

\author{%
Jennifer Rittenhouse West}
\email{jrwest@slac.stanford.edu}
\affiliation{Department of Physics and Astronomy, University of California, Irvine, CA 92697, USA} 
\affiliation{SLAC National Accelerator Laboratory, Stanford University, Stanford, California 94309, USA}

\date{\today}

\begin{abstract}
Under the assumption that the current epoch of the Universe is not special, i.e. is not the final state of a long history of processes in particle physics, the cosmological fate of $SU(3)_C \times U(1)_{\rm EM}$ is investigated.  Spontaneous symmetry breaking of $U(1)_{\rm EM}$ at the temperature of the Universe today is carried out.  The charged scalar field $\phi_{\rm EM}$ which breaks the symmetry is found to be ruled out for the charge of the electron, $q=e$.  Scalar fields with millicharges are viable and limits on their masses and charges are found to be $q\lesssim10^{-3}e$ and $m_{\phi_{\rm EM}}\lesssim10^{-5} \rm eV$.  Furthermore, it is possible that $U(1)_{\rm EM}$ has already been broken at temperatures higher than $T=2.7K$ given the nonzero limits on the mass of the photon.  A photon mass of $m_{\gamma}=10^{-18} \rm eV$, the current upper limit, is found to require a spontaneous symmetry breaking scalar mass of $m_{\phi_{\rm EM}}\sim 10^{-13} \rm eV$ with charge $q=10^{-6}e$, well within the allowed parameter space of the model.  Finally, the cosmological fate of the strong interaction is studied.  $SU(3)_C$ is tested for complementarity in which the confinement phase of QCD $+$ colored scalars is equivalent to a spontaneously broken $SU(3)$ gauge theory.  If complementarity is not applicable, $SU(3)_C$ has multiple symmetry breaking paths with various final symmetry structures.  The stability of the colored vacuum at finite temperature in this scenario is nonperturbative and a definitive statement on the fate of $SU(3)_C$ is left open.  Cosmological implications for the metastability of the vacua - electromagnetic, color and electroweak - are discussed.

\vspace{11mm}
\end{abstract}

\maketitle

\section{Introduction}
\label{sec:intro}
Symmetries of Nature are deeply connected to particles and their interactions.  The fundamental symmetry structure of the Universe has changed at least once over the past $13.8 \times 10^9$ years.  The early Universe combined the weak and electromagnetic interactions, a symmetry that was broken by the Higgs field and is described by the Standard Model (SM) of particle physics.  The SM contains all currently known particles and interactions, with the exception of neutrino masses, as well as the spontaneous symmetry breaking (SSB) mechanism of the Higgs field.

Grand unified theories (GUTs) unify the strong and electroweak sectors of particle physics into larger symmetry groups which are typically broken down to the SM by new scalar fields at earlier times and higher temperatures \cite{Buras:1977yy}.  They are highly motivated by physics beyond the SM, notably dark matter and quantum gravity, and if realized in Nature would extend the symmetry breaking pattern of the past.  

It may be of interest to study symmetry breaking in the future.  If the current age and temperature of the Universe are not special with respect to symmetries, then as the previous group structure of the Universe was broken at least once, so may the current structure be broken one or more times.  In this case the fate of the Universe may be determined by studying the possible symmetry breaking paths and the effects thereof.

The particle physics framework of the $2.7~K$ Universe is the gauge group structure $SU(3)_C \times U(1)_{\rm EM}$.  The remarkable success of both QED and QCD in predicting particle properties, decays and interactions gives compelling evidence that these local symmetries hold today.  At earlier times, i.e. at temperatures greater than $T \sim 100~ \rm GeV$, electroweak symmetry breaking had not yet occurred and the larger group structure of the SM, $SU(3)_C \times SU(2)_L \times U(1)_Y$, held.  The hypothesis of spontaneous symmetry breaking of the SM via the scalar Higgs field was confirmed in 2012 in a stunning achievement of experimental collider physics \cite{Aad:2012tfa,Chatrchyan:2012xdj}.  That discovery, a proof of existence in a sense, allows for the question of future SSB.  For SSB to occur, new scalar fields with color and/or electromagnetic charge are needed.  The shape of the the scalar potential must be such that the proposed $SU(3)_C \times U(1)_{\rm EM}$ vacuum is metastable.  This work investigates whether the current group structure is truly the final symmetry state of the Universe.

The cosmological fate of $SU(3)_C \times U(1)_{\rm EM}$ may, under highly specific conditions, affect the fate of the Universe.  If $U(1)_{\rm EM}$ is spontaneously broken by an electromagnetically charged scalar field, the cosmic microwave background (CMB) photons gain mass.  The photon mass, as shown in Section \ref{sec:em_breaker}, is dependent upon the vacuum expectation value (vev) of the scalar field,
\bea
m_{\gamma} = \sqrt{2}q v,
\eea
where $q$ is the charge of the scalar field in units of electron charge and $v$ is the vev.

If the acceleration of the expansion of the Universe is caused by a cosmological constant, $\Lambda$, this gain in mass (regardless of the value of the vev) will have no effect on the fate of the Universe under the usual assumption of a Friedmann-Lema\^{i}tre-Robertson-Walker (FLRW) Universe.  In the homogenous, isotropic and flat FLRW Universe the total energy density, having previously passed through phases of radiation domination followed by matter domination, has recently entered the cosmological constant dominated phase.  During this phase, nothing can overtake the effect of $\Lambda$'s energy density on the expansion rate and the expansion will be eternal and accelerated \cite{Carroll:2000fy}.  

However, if the acceleration is not caused by a cosmological constant and instead dark energy evolves in the future in such a way that its energy density parameter varies as the scale factor $a^y(t)$ with $y< -3$, the CMB photons' gain in mass could be important.  Radiation is the only currently known type of energy density that evolves in the necessary way, as $a^{-4}(t)$.  If the dark energy were to be modeled by a scalar field (or fields) which decays in the future to radiation then the breaking of $U(1)_{\rm EM}$ could affect the fate of the Universe given a large enough value of $m_{\gamma}$.  The expansion rate could slow down or even reverse.  

The 2018 cosmological parameters from the \textit{Planck} satellite strongly favor a small positive cosmological constant today \cite{Aghanim:2018eyx}.  There is currently no reason to believe the evolution outlined above would occur, however, the cause of the acceleration of the expansion is unknown and some kind of evolution in time is plausible.

For the SSB potential considered in this work, the allowed masses for an electromagnetically charged scalar field are quite small and must be millicharged in order to be viable.  More complicated scalar potentials, e.g. composed of scalar fields carrying both electromagnetic and color charge, or the use of a non-SSB mechanism (e.g. radiative symmetry breaking \cite{PhysRevD.7.1888}) could affect this conclusion.  Previous studies of a charged Higgs boson related to the SM Higgs at finite temperatures did not support a SSB, a fact relayed to the author after submitting an earlier version of this work \cite{Primsher:1980abc}.

It would be interesting to allow for higher mass millicharged fields as these are excellent dark matter candidates, e.g. \cite{Barkana:2018lgd}.  New experiments such as the Light Dark Matter Experiment (LDMX) \cite{Berlin:2018bsc}, MilliQuan \cite{Ball:2016zrp}, NA64 \cite{Gninenko:2018ter} and SHiP \cite{SHiP:2018yqc} propose to detect $\sim1$ MeV to $\sim10$ GeV particles with charges from $10^{-1}e$ to $10^{-4}e$.  Sub-MeV millicharged particle tabletop detectors are currently in development as well.  The millicharged particles discussed in this work are too light for these direct detection experiments but they may be of interest for next generation experiments.  There is a possibility of pushing to higher masses while retaining millicharges, discussed in Section \ref{sec:highT}.

\section{Stability of the Vacua}
\label{sec:vacuum}
The implicit assumption is that the electroweak vacuum is stable and the SM Higgs vacuum expectation value of $246 ~ \rm GeV$ is the true vacuum.  This may not be true.  More precise measurements of the Higgs coupling to the top quark are needed to determine the stability of the electroweak vacuum, with consequences of metastability outlined in the 1980s \cite{TurnerWilczek}.  A state-of-the-art calculation \cite{PhysRevLett.119.211801} suggests that we are in a metastable electroweak vacuum with a lifetime of 

\bea
\tau =10^{561^{+817}_{-270}} ~\rm years, 
\eea
with the given uncertainties due only to the top quark mass (other SM parameter measurement uncertainties contribute but the top quark mass dominates).  This result may settle into absolute stability or shorter lifetime metastability by physics beyond the SM, including a theory of quantum gravity \cite{Branchina:2013jra,2017arXiv170801138B}.  The necessary precision on the top quark mass for a $3\sigma$ metastability confirmation is $\Delta m_t < 250 ~ \rm MeV$ \cite{2017arXiv170708124A}.  With the current uncertainty of the top quark mass from direct measurements, $m_t=173.21 \pm 0.51 \pm 0.71 ~ \rm GeV$ \cite{Patrignani:2016xqp}, this question will likely not be answered with additional Large Hadron Collider (LHC) Run II results.  Uncertainties of less than $200 ~ \rm MeV$ may be accessible when the high luminosity (HL-LHC) upgrade is complete and the full dataset taken \cite{Husemann:2017eka}.

The stability of a QCD-QED vacuum is a separate question.  The vacuum metastability investigated in this case is due to a colored scalar field or fields with a nonzero vev and/or an electromagnetically charged scalar field with a nonzero vev.  The presence of a new scalar charged under $SU(2)_W \times U(1)_Y$ - as any field with $q \neq 0$ must be - affects the shape of the Higgs potential at high energies and therefore may have an effect on electroweak vacuum stability.  

If it were possible to rule out the existence of a new electromagnetically charged scalar field, the electroweak vacuum stability question would remain dependent upon future precision measurements of the top quark mass as well as any new physics effects.  The sub-eV mass millicharged scalar fields used in the model presented here would have very little effect on the shape of the Higgs potential but with higher mass millicharged scalar fields (discussed in Section \ref{sec:highT}) this could change.

\section{$U(1)_{\rm EM}$ spontaneous symmetry breaking}
\label{sec:em_breaker}
In order to break the $U(1)_{\rm EM}$ gauge symmetry, an electromagnetically charged scalar field $\phi_{\rm EM}$ is introduced (the subscript will be dropped for clarity).  It is a color singlet with charge $q$ under $U(1)_{\rm EM}$ and gives rise to a new scalar section of the QED Lagrangian, 
\bea
\mathscr{L}_{QED} \supset -\frac{1}{4} F^{\mu\nu}F_{\mu\nu} + D^{\mu}\phi^{*}D_{\mu}\phi  - V(\phi),
\eea
with covariant derivative $D_{\nu} = \partial_\nu + i q A_\nu$ and field transformations under $U(1)_{\rm EM}$

\bea
\phi \rightarrow e^{i\alpha} \phi, ~ A_\nu \rightarrow A_\nu - \frac{1}{e} \partial_\nu \alpha.
\eea
The scalar potential is given by
\bea \label{eqn:tree}
V(\phi)= -\frac{\mu^2}{2} |\phi|^2 + \frac{\lambda}{4!}(\phi^{*}\phi)^2
\eea

where the field $\phi$ gains a vev for the choice of mass parameter $\mu^2 > 0$.  Labeling the minimum of the potential $v_e$,

\bea
v_e^2 \equiv \langle \phi \rangle^2 = \frac{6\mu^2}{\lambda}
\eea
the Lagrangian is expanded about the minimum.  The complex scalar field may be written as $\phi = v_e + \frac{1}{\sqrt{2}}(\phi_1 + i \phi_2)$ to yield
\bea
V(\phi)= -\frac{3}{2\lambda} \mu^4 + \mu^2 \phi_1^2 + \mathscr{O}(\phi_i^3)
\eea

One of the scalar fields gains mass $m_{\phi} = \mu$ and the other is the massless pseudo-Goldstone boson which provides the longitudinal polarization of the now massive photon.

The photon gains its mass via the kinetic energy term of $\mathscr{L}$, 
\bea
m_A^2 = 2q^2 v_e^2.
\eea





\subsection{Electromagnetic scalar potential in the $2.7~K$ Universe}
In order to study the effects of new scalar fields in the current epoch, we calculate the scalar potential at the temperature of the cosmic background radiation today.  Computing $V(\phi,T)$ for $T \approx 10^{-4}~\rm eV$ gives an estimate of the effects in terms of relationships between the parameters of the model.

It is important to note that the finite temperature field theory equations assume both equilibrium conditions and homogeneity of the medium.  To accommodate the non-equilibrium conditions - the CMB has a thermal distribution but is not in equilibrium due to the expansion of the Universe - a time slice at $T=2.7~\rm K$ is used.  Equilibrium is assumed for this moment in time.  The assumption of homogeneity in the Universe is length scale dependent.  On the largest length scales both homogeneity and isotropy appear to hold.  This work is concerned with such cosmological scales.

Closely following the treatment of Quir\'{o}s 1999 \cite{Quiros:1999jp} and Coleman and Weinberg 1973 \cite{PhysRevD.7.1888}, the finite temperature potential in terms of the constant background field $\phi_e$ is given by 
\bea
\hspace{-3mm}V(\phi_e, T) = V_{0}(\phi_e) + V_{1}(\phi_e, 0) + V_{1T}(\phi_e, T) \ ,
\label{eff_pot}
\eea

where the first term is the zero temperature classical potential as in \eqref{eqn:tree},
the second term is the zero temperature Coleman-Weinberg correction to one-loop order and the final term is the finite temperature contribution, also calculated to one-loop.

Both zero temperature and finite temperature loop calculations include contributions from all relevant particles coupled to the scalar field.  The gauge boson of the $U(1)_{\rm EM}$ gauge group, the fermions charged under it, and the scalar field itself all may run in the loop.  Fermions will not be relevant for the temperatures and densities considered here due to the baryon-to-photon ratio data as given by Big Bang Nucleosynthesis (BBN) \cite{PhysRevD.98.030001}

\bea
\begin{aligned}
5.8 \times 10 ^ { - 10 } \leq n _ { \mathrm { b } } / n _ { \gamma } \leq 6.6 \times 10 ^ { - 10 } ( 95 \% \mathrm {CL}),
\end{aligned}
\eea

\noindent however photons and $\phi$ both contribute to the effective potential.

The 1-loop $T=0$ contributions are given, using $\overline{MS}$ renormalization counter terms with a cut-off regularization and the assumption that the minimum and the scalar mass do not change with respect to their tree level values, that is
\bea
\begin{split}
\frac{d(V_1+V_1^{c.t.})}{d\phi_e}\biggr\vert_{\phi_e = v_e} = 0 \\
\frac{d^2(V_1 + V_1^{c.t.})}{d(\phi_e)^2}\biggr\vert_{\phi_e = v_e} = 0, \\
\end{split}
\eea

to be

\bea
\begin{aligned}
V_1(\phi_e)= \frac{1}{64\pi^2}\sum_i n_i \{m_i^4(\phi_e) (\log{\frac{m_i^2(\phi_e)}{m_i^2(v_e)}} - \frac{3}{2}) ~+ \\
 2 m_i^2(v_e)m_i^2(\phi_e) \}.
\end{aligned}
\eea

Here $i=\gamma, \phi$ and $n_i$ the degrees of freedom with $n_\gamma=3$ for the newly massive photon and $n_\phi=1$ for the scalar field.

The contributions to the thermal effective potential to 1-loop order are given by

\bea
V_{1T}\!=\! \sum_i\!\frac{n_i}{2 \pi^2 \beta^4}\! \int_0^\infty{\!\!\!\!\! dx ~ x^2 \log\!\left(\!1\!-\!e^{-\sqrt{x^2+\beta^2 m_i^2(\phi_e)}}\right)}\!.
\eea

The high temperature expansion cannot be used in the case of the $T=2.7K$ Universe and an analytic solution to the temperature dependent integral does not exist.  However, a numerical solution is possible under the conditions outlined in the following subsection.  

\subsection{Spontaneous symmetry breaking conditions}
Any future spontaneous symmetry breaking will depend upon the sign of the quadratic coefficient in the effective potential, $\frac{d^2 V}{d\phi_e^2}$ evaluated at $\phi_e\!=\!0$.  Two constraints must be satisfied.  First, that there has been no SSB until today.  This stability condition becomes
\bea
\frac{d^2 V}{d\phi_e^2}\biggr\vert_{\phi_e=0, ~ T\geq 2.7K} \geq 0.
\eea
Second, that a SSB may occur in the future, and let us take the furthest future possible in temperature, i.e. $T=0$, 

\bea
\frac{d^2 V}{d\phi_e^2}\biggr\vert_{\phi_e=0, T= 0} < 0.
\eea

When this second derivative is negative for temperatures $T< 2.7K$, with the additional requirement that $\lambda > 0$, SSB will occur.  Setting these constraints on $V(\phi_e,T)$ allows for a numerical evaluation of the integrals for any value of $\mu^2$ and $\lambda$, as the derivatives may be taken prior to integration.  The equation to be constrained is

\bea
\frac{d^2 V}{d\phi_e^2}\!\!= \!-\mu^2\!+\!\frac{3q^4\mu^2}{2\pi^2\lambda}\!-\!\frac{\lambda \mu^2}{64\pi^2}\!+\!\frac{q^2T^2}{2} \!+\! \lambda T^2 
f\!\left(\!\frac{\mu^2}{T^2}\!\right)
\label{eq:derivative}
\eea

where the function $f$ is the second derivative of the thermal bosonic function in \cite{Quiros:1999jp}, evaluated for the scalar boson at $\phi_e\!=\!0$ with the quadratic temperature dependence and the quartic coupling factored out.  It is given by

\bea
\!\!\!f\!\left(\!\frac{\mu^2}{T^2}\!\right)\!\!=\!\!\frac{1}{2\pi^2}\!\int_0^{\infty}{\!\!\!\!\!dx \frac{x^2~e^{-\sqrt{x^2-\frac{\mu^2}{T^2}}}}{\left( 1-e^{-\sqrt{x^2-\frac{\mu^2}{T^2}}}\right) \sqrt{x^2-\frac{\mu^2}{T^2}}}}.
\eea

\subsection{Charge $\frac{q}{e}=1$ Scalar Fields}
For a charge equal to the electron charge, satisfying both SSB conditions with $0\!<\!\lambda \!\!< \!4\pi$ requires the allowed masses of $\phi$ to be too light, much less than the mass of the electron.  Such particles would have been produced, for example, at the Large Electron-Positron Collider (LEP) in great quantities and they were not detected, thus ruling out a SSB for $U(1)_{\rm EM}$ with a $q=e$  field.  More concretely, the light masses found here are neatly excluded by the SLAC Anomalous Single Photon (SLAC ASP) search which ruled out $q>0.08e$ for $m_{\rm MCP}\lesssim 10$ GeV, a result holding for any weakly interacting millicharged particle \cite{Essig:2013lka}.

The stability constraint is satisfied for any $\mu^2\!<\!0$ and any charge $q$ as well as for $\mu^2>0$ with ranges of allowed $q$ and $\lambda$, so a heavier electrically charged scalar field with the tree level potential in \eqref{eqn:tree} is possible in Nature.  However such a scalar could not be the source of spontaneous symmetry breaking for $U(1)_{\rm EM}$ and could not give the photon a mass.

\subsection{Millicharged (Minicharged) Scalar Fields}
Millicharged scalar fields can spontaneously break $U(1)_{\rm EM}$.  Millicharge describes any charge less than that of the electron (i.e. not exclusively $10^{-3}e$), although it may also mean any charge less than that of the down quark, $q<\frac{1}{3}e$.  The more accurate (but less used) term is minicharged particles.  

The key to a SSB in the finite temperature Universe in this case is to choose parameters that yield an unstable potential at $T=0$ which gives a SSB in the future and is easily accomplished.  The first term in Eqn. \eqref{eq:derivative} is negative and dominates the other terms even with a $q=e$ choice for the charge.  There are no constraints on the mass in this case, as stated in the previous section.

Next, turn on the finite temperature loop contributions and find parameter ranges that create an overall positive coefficient for the $\phi^2$ terms.  The finite temperature pieces are small.  The first term arises from the photon running in the loop of the background scalar field.  With $T=2.7K$ or $\approx10^{-4}$ eV and a millicharge even as large as $q=10^{-2}e$ it is of order $10^{-12}$.  The next term is the scalar boson running in the loop and is bounded by $\lesssim 10^{-8}$ eV for the ranges of masses and charges tested here.  The constraint equation becomes

\bea
\abs{-\mu^2+\frac{3q^4\mu^2}{2\pi^2\lambda}-\frac{\lambda\mu^2}{64\pi^2}} \lesssim 10^{-n}
\eea
where $10^{-n}$ is defined to be the size of the finite temperature terms.  The term inside of the absolute value is negative, its magnitude must be less than that of the positive finite temperature terms.  For $\lambda=1$ and $q=10^{-3}e$, this gives $m_{\phi_{\rm EM}}\lesssim10^{-5}$ eV with a finite temperature term of order $10^{-8}$.  Smaller masses and smaller charges are also viable.

Upon inspection of Eqn. \ref{eq:derivative} it appears that small enough values of the quartic coupling $\lambda$ could open up a larger parameter space but in fact this is not so.  A very small $\lambda$, $\lambda\ll1$, can force a positive overall $\phi^2$ coefficient.  The condition $\lambda<\frac{3q^2}{2\pi^2}$ forces a stable scalar potential.  For $q=10^{-3}e$, $\lambda<10^{-12}$ satisfies this with no constraint on the mass of the scalar.   On the other hand, the now $2$ conditions $\lambda>\frac{3q^4}{2\pi^2}$ and

\bea
\abs{\mu^2\left( \frac{3q^4}{2\pi^2 \lambda}-1 \right)}>\frac{q^2T^2}{2}
\eea
force a metastable scalar potential.  For the previously considered $q\sim10^{-3}e$ and $m_{\phi_{\rm EM}}\sim10^{-5}$ eV, $\lambda \sim 10^{-13}$ accomplishes the task.  The problem is managing a transition between these two states, either by varying $\lambda$ with temperature/time or by some other means.  It does not seem possible to do this.

Astrophysical bounds on millicharged particles from stellar cooling constraints set limits on $m \lesssim$ keV masses requiring charges $q \lesssim 10^{-15}e$ \cite{2017JHEP...02..033H}.  The very light masses considered here allow for charges within these upper bounds.  For $q=10^{-15}e$, $\lambda=1$, the necessary mass is of the order $m_{\phi_{\rm EM}}\sim10^{-5}$ eV.  The final term in Eqn. \ref{eq:derivative} is independent of the coupling $q$, thus the scalar mass does not change much from the previous value.


\subsection{$U(1)_{\rm EM}$ Breaking in the $T>2.7K$ Universe}
\label{sec:highT}

It may be that the $U(1)_{\rm EM}$ has already been broken by a millicharged scalar field.  The accepted 2018 limits on the mass of the photon, $m_\gamma \lesssim10^{-18} \rm eV$, come from magnetohydrodynamic studies of the solar wind \cite{PhysRevD.98.030001}.  Tighter limits are given by studies of the galactic magnetic field but depend critically on assumptions that may not hold, e.g. the applicability of the virial theorem.  However, for the sake of completeness, the tightest limits of $m_\gamma \lesssim 10^{-27} \rm eV$ are also discussed here.

The accepted limit of $m_\gamma = 10^{-18} \rm eV$ requires a mass for the millicharged field of $m_{\phi_{\rm EM}}= 3\times10^{-13}$ eV for a charge of $q=10^{-6}e$ and $\lambda=1$, well within the bounds found here.  For $q=10^{-3}e$ and $q=10^{-9}e$, the necessary masses are $m_\phi$ of $10^{-16}$ eV and $10^{-10}$ eV respectively.

The smaller $m_\gamma \lesssim 10^{-27} \rm eV$ upper bound on the mass of the photon requires $m_\phi = 3\times 10^{-22}$ with the same charge and quartic coupling value, also within the limits of the model.

Recent work \cite{Reece:2018zvv} suggests that determining whether the Standard Model photon is exactly massless or not is of interest to light mass dark photon model builders as a strict $m_\gamma=0$ rules out some of this model parameter space.

In order to accomplish this, it must be that they gain mass via the St{\"u}ckelberg mechanism (\cite{Ruegg:2003ps} and references within) and not a Higgs mechanism.  The SSB mechanism here is able to give photons a mass without putting any restrictions on light dark photon models.

\section{$SU(3)_C$ symmetry breaking - Complementarity}
\label{sec:colorcomp}

Breaking the symmetry of QCD is not straightforward.  The strong coupling $\alpha_s$ is nonperturbative at $2.7~ \rm K$.  The effects of SSB in the $2.7 ~ \rm K$ Universe require finite temperature field theory calculations.  Lattice QCD is needed to calculate the nonperturbative corrections to the effective potential of a new color charged scalar field, $\phi_c^a$, at temperatures below $\Lambda_{QCD}$.

Complementarity between the confined phase of QCD $+$ $\phi_c^a$ and the broken symmetry phase of $SU(3)_C$ may be able to eliminate the need for a SSB to investigate the fate of $SU(3)_C$.  A test for the applicability of complementarity has been proposed by Georgi \cite{Georgi:2016qbt}.  According to that work, the structure of the heavy stable particles of the confined phase must match that of the broken symmetry phase in order for complementarity to hold.

In order to make use of complementarity, however, a continuously varying parameter must take the confined phase of QCD $+$ a colored scalar field to the spontaneously broken $SU(3)_C$.  The mass parameter of a colored scalar field (or fields) is a natural choice as it mirrors the process of $U(1)_{\rm EM}$.  The finite temperature effective potential calculations for colored scalars require the use of lattice QCD.  There is a possibility of $3$ colored scalar fields being able to manage the transition [H. Georgi, personal communication] but the nonperturbative calculations are far beyond the scope of this work.  It may be of interest to note that $3$ colored scalar fields are $1$ more than is necessary for a SSB of $SU(3)_C$ down to no gauge structure at all. \\

\section{$SU(3)_C$ Symmetry Breaking - SSB}
\label{sec:break}

With the applicability of complementarity unclear, the possibility of a carrying out a symmetry breaking of $SU(3)_C$ remains.  Both the adjoint and fundamental representations of $SU(3)_C$ are a priori viable as the masses of the gluons are unconstrained in the future.  When restricting to SSB, at least $1$ colored scalar field in the fundamental representation is added to the theory of QCD and multiple final states of gauge symmetries are possible.   For SSB of $SU(3)_C$ to end with no local symmetries, 2 new colored scalar fields are needed.  The other possible final symmetry states, a gauged $SU(2)$ or $U(1)$, are realized with one new colored scalar field.  An example of a final $SU(2)$ symmetry state is sketched out next.

For $SU(3)_C$ to be spontaneously broken to an $SU(2)$, a colored scalar triplet, $\phi_c^a$ is proposed.  This new Higgs field has a $({\bf 3},0)$ assignment under $SU(3)_C \times U(1)_{\rm EM}$ and removes one rank from the strong force gauge group, leaving the structure $SU(2)_{BC} \times U(1)_{\rm EM}$.  Here BC stands for "broken color."  The minimal effective potential at tree level is 
\bea
V_0(\phi_c)=-m_c^2 \phi_c^{\dagger}\phi_c + \lambda_c | \phi_c^{\dagger}\phi_c  |^2.
\eea
with color indices suppressed.

On purely dimensional analysis grounds it may be argued that any contribution to $\frac{d^2 V}{d\phi_c^2}$ from nonperturbative corrections would be of the order $\Lambda_{QCD}^2$.  In order for SSB to occur, it must be that the quantity 
\bea
\frac{d^2V(\phi_c)}{d\phi_c^2}\biggr\vert_{\phi_c=0} =-m_c^2 + \Lambda_{QCD}^2 + f_c\left(\!\frac{m_c^2}{T^2}\!\right)
\eea
transitions from a positive number to a negative number.  The finite temperature pieces $f_c\left(\!\frac{m_c^2}{T^2}\!\right)$ are suppressed by the lightest QCD particles, the $\sim \!\!~ \!100 \rm ~ MeV$ pions.  This implies that $m_c^2$ and $\Lambda_{QCD}^2$ are very nearly the same, i.e. the mass of the colored scalar particle would be on the order of hundreds of MeV.  Colored scalars in this mass range could bind strongly to single quarks, forming a pion-like system of spin $\frac{1}{2}$ rather than spin $0$.  Such mesons would have been detected long ago.  In particular, for $\rm e^+ e^-$ colliders, the $R$ ratio of the hadronic cross section to the muonic cross section 

\bea
R = \frac { \sigma ^ { ( 0 ) } \left( e ^ { + } e ^ { - } \rightarrow \text { hadrons } \right) } { \sigma ^ { ( 0 ) } \left( e ^ { + } e ^ { - } \rightarrow \mu ^ { + } \mu ^ { - } \right) }
\eea
has been extremely well measured \cite{Eidelman:1995ny} and colored scalars of this mass range are ruled out.   

A precise calculation is needed for a definitive statement. As mentioned previously, the effects of new colored scalars are highly nontrivial and are not explored further here.

\section{Conclusions}
\label{sec:conclu}

The cosmological fate of the electromagnetic and strong interactions have been investigated under the assumption that we currently live in an intermediate rather than final stage of the symmetries, charges and interactions of particle physics.  

It is found that $U(1)_{\rm EM}$ will remain an infinite range interaction forever for the case of a scalar field with the charge of the electron which is a singlet under $SU(3)_C$.  

The more interesting case of a millicharged scalar field, still a singlet under $SU(3)_{C}$, is viable within a specific range of masses and millicharges.  It is capable of spontaneously breaking the symmetry of $U(1)_{\rm EM}$ and may also be a dark matter candidate \cite{West:2019boa}.  This scenario gives a mass to the photon which is light enough that it would not change the cosmological expansion rate $H(z)$ for $z<0$ and would likely not affect the fate of the Universe.

$U(1)_{\rm EM}$ may already have been spontaneously broken by a scalar of mass $m_{\phi_{\rm EM}}\lesssim 10^{-13}$ eV for a photon mass equal to the current upper bound of $m_{\gamma} \sim10^{-18}$ eV and a charge of $q=10^{-6}e$.  If the astrophysical limits from the galactic magnetic field studies on $m_{\gamma}$ hold, this upper bound becomes $m_{\gamma}\sim 10^{-27}$ and the mass of the millicharged field $m_{\phi_{\rm EM}}\lesssim10^{-22}$ again for microcharges $q=10^{-6}e$.  The higher temperatures of the earlier Universe appear to lift the low mass constraint on $m_\phi$ and this will be explored in a followup paper \cite{West:2019boa}.  This would push $\phi_{\rm EM}$ into the realm of detection with the dark matter experiments discussed in Section \ref{sec:intro}.

The fate of $SU(3)_C$ is likely to remain unbroken but is as yet unknown.  

If it emerges that $U(1)_{\rm EM}$ was broken at higher temperatures in the early Universe or will be broken in the future, there may be some aesthetic appeal to either breaking the one remaining symmetry of $SU(3)_C$ or forbidding it from being broken.  Should this be the case, future work would include testing Georgi's complementarity principle in earnest with collaborators in the QCD community.  The successful implementation of complementarity along with a broken $U(1)_{\rm EM}$ would yield a final state of the Universe with no local symmetries at all, satisfying an evolution from early Universe higher symmetry structures to none at all at late times.  
\\

\subsection*{Acknowledgments}

I thank Tim Tait, Howard Georgi and Michael Ratz for useful discussions on an earlier version of this paper.  

I am grateful to the theory group at SLAC National Accelerator Laboratory as well as the University of California, Santa Cruz for their insightful questions and comments on millicharged fields at the SLAC theory seminar where this work was presented.  I am particularly grateful to Howard Haber for his question on applying this model to the current limits on the mass of the photon as this opened up a relevant new avenue for me to pursue.

This work was completed at SLAC National Accelerator Laboratory and I am indebted to the theory group for their kind hospitality, especially Stan Brodsky, Michael Peskin and Tom Rizzo.  

This research was partially supported by the GAANN Federal Fellowship.

\bibliography{break}
\bibliographystyle{unsrt}

\end{document}